\documentstyle[prl,aps,psfig]{revtex}
\begin{document}
\twocolumn[\hsize\textwidth\columnwidth\hsize
           \csname @twocolumnfalse\endcsname
\title{Budd-Vannimenus theorem for superconductors}
\author{Pavel Lipavsk\'y$^1$, Klaus Morawetz$^{2,3}$, 
Jan Kol\'a\v cek$^1$, 
Ji{\v r}{\'\i} J. Mare{\v s}$^1$, Ernst Helmut Brandt$^4$}
\address{$^1$Institute of Physics, Academy of Sciences, 
Cukrovarnick\'a 10, 16253 Prague 6, Czech Republic\\
$^2$Max-Planck-Institute for the Physics of Complex
Systems, Noethnitzer Str. 38, 01187 Dresden, Germany\\
$^3$Institute of Physics, Technical University Chemnitz, 09107 Chemnitz\\
$^4$Max-Planck-Institut f\"ur Metallforschung,
         D-70506 Stuttgart, Germany}
\maketitle
\begin{abstract}
The Budd-Vannimenus theorem is modified to the surface of 
superconductors in the Meissner state. This identity links the 
surface value of the electrostatic potential to the free energy in 
the bulk which allows one to evaluate the observed potential 
without the explicit solution of the charge profile at the surface. 
\end{abstract}
    \vskip2pc]

The Budd-Vannimenus theorem \cite{BV73} links the value of 
the electrostatic potential at the surface of a metal to the 
energy per electron in the bulk. This simple consequence of 
the Feynman-Hellmann theorem turned out to be very useful in 
studies of metal surfaces \cite{KW96} as it enables one to 
circumvent demanding calculations of the charge distribution 
at the surface. We will use this approach to evaluate the 
surface potential of superconductors within the 
Ginzburg-Landau (GL) theory. 

Perhaps it is useful to start from a historical review. Within 
the pre-London approaches based on the ideal charged liquid 
\cite{B37}, the electrostatic potential was assumed to balance 
the Lorentz and the inertial forces acting on currents. Since 
the current $j=env$ flows along the surface and its amplitude 
falls off exponentially from the surface into the bulk, the 
electrostatic Bernoulli--type potential, $e\varphi=-
{1\over 2}mv^2$, falls off on the scale of the London penetration 
depth. From the Poisson equation, $\rho=-\epsilon_0\nabla^2 
\varphi$, one finds that this corresponds to a charge 
accumulated in the layer penetrated by the magnetic field.  
Charge neutrality is maintained by the opposite 
surface charge. This picture has been confirmed by the London 
theory \cite{L50}.

The electrostatic potential equals the Bernoulli potential only at 
zero temperature when all electrons are in the superconducting 
condensate. At finite temperatures, a part $n_n$ of electrons 
remains in the normal state, while the rest $n_s=n-n_n$ 
contributes to the supercurrent, $j=en_sv$. As a consequence, 
the electrostatic potential reduces to $e\varphi=-{n_s\over n}
{1\over 2}mv^2$ as shown by van~Vijfeijken and Staas 
\cite{VS64}. Jakeman and Pike \cite{JP67} recovered this 
result from the time-dependent GL theory. Moreover, their 
approach describes the surface charge as a space charge 
localized on the Thomas-Fermi screening length.

The surface charge and the charge accumulated at the London 
penetration depth form a surface dipole due to which the 
electrostatic potential in the bulk differs from the potential 
out of the superconductor. This voltage difference, 
which like the resistance of the Hall effect is perpendicular to the 
current, is not accessible by contact measurements but it was 
observed by a contact-less capacitive pickup 
\cite{BK68,BM68}. 

The first measurements \cite{BK68,BM68} were not sufficiently 
accurate and created expectations that the surface voltage 
includes traces of the pairing mechanism. Rickayzen \cite{R69} 
used the thermodynamic approach and showed that the 
potential indeed has a pairing contribution, $e\varphi=-
{n_s\over n}{1\over 2}mv^2-4{n_n\over n}{\partial\ln T_c\over
\partial\ln n}{1\over 2}mv^2$. The first term results from the 
Lorentz and inertial forces, while the second one reflects the 
pairing. The pairing term dominates close to the critical temperature 
$T_c$. If $\varphi$ would be experimentally accessible, one 
could deduce from it the density dependence of $T_c$.
The highly accurate measurement by Morris and Brown 
\cite{MB71}, however, chilled all the expectations. The 
electrostatic potential revealed no pairing contribution. 

The disagreement between the theory and experiment remained 
unexplained for a long time and the question of the
charge transfer in superconductors was left aside 
till the discovery of the high-$T_c$ materials. It was 
predicted\cite{M89,KK92} that in layered materials the
superconducting transition induces a charge transfer from 
Cu0$_2$ planes to charge reservoirs. This transfer caused
merely by the pairing mechanism has been experimentally
confirmed by the positron annihilation\cite{positron}, the
$x$-ray absorption spectroscopy\cite{xray}, and the nuclear
magnetic resonance\cite{KNM01}.

Apparently, there are two groups of contradictory experimental
results. The pairing contribution is absent in the surface 
potential but it is observed in the bulk. As it was indicated 
recently \cite{LKMM02}, there is an additional surface dipole 
which cancels the pairing contribution to surface potential 
seen in capacitive measurements.

The approach presented in Ref.~\cite{LKMM02} employs the 
Budd-Vannimenus theorem, however, it has the same 
limitation as Rickayzen's formula -- the magnetic field is 
described within the London's theory and it has to be weak not 
to perturb the density of the condensate. 
In contrast, the measurement of Morris and Brown \cite{MB71} 
explores the entire range of the magnetic fields from weak 
fields up to the critical value. They found that the potential 
equals the magnetic pressure $\varphi={1\over \rho_{\rm lat}}
{1\over 2\mu_0}B^2$, i.e., it is independent of the temperature 
and the used material enters merely via the charge density
$\rho_{\rm lat}$ of its lattice, for both type-I and II 
superconductors. 

Here we modify the Budd-Vannimenus theorem so that it is 
consistent with the GL theory. Its validity is not restricted 
to weak magnetic fields. We will show that the surface 
potential is given by
\begin{equation}
\varphi_0={f_{\rm el}\over \rho_{\rm lat}},
\label{e7}
\end{equation}
where $f_{\rm el}$ is the surface value of the GL free energy 
(\ref{e2}). With the help of the modified Budd-Vannimenus 
theorem one can explain the experimental result of Morris and 
Brown under very general conditions.

The recent implementation of the GL theory has changed the 
picture of the surface charge. It is not localized on the 
scale of the Thomas-Fermi 
screening length as claimed by Jakeman and Pike, but it 
varies on scales of the GL coherence length \cite{KLB02}. As pointed 
out in Ref.~\cite{YBPK01}, if the electrostatic potential is a 
local functional of the GL wave function, $\varphi[\psi]$, the GL 
boundary condition, $\nabla\psi=0$ in the direction normal 
to the surface, implies the zero electric field at the surface, 
$E=-\nabla\varphi={\partial\varphi\over\partial\psi}\nabla\psi=0$. 
Accordingly, no additional surface charge is needed to 
maintain the charge neutrality. 

Since the surface charge is absent, it seems that the surface 
dipole on the Thomas-Fermi screening length is absent, too. 
There has to be yet another surface dipole, however. The pairing 
correlation is weaker on the surface than in the bulk, what results 
in forces pulling the Cooper pairs inside. Such forces are always
balanced by the electrostatic field. The understanding of this 
effect will require microscopic studies which are not yet feasible. 
We can merely speculate that the surface dipole is somehow linked to 
the gap profile at the surface \cite{G66}, i.e., it is localized on 
the BCS coherence length.

\begin{figure}[h]  
\psfig{figure=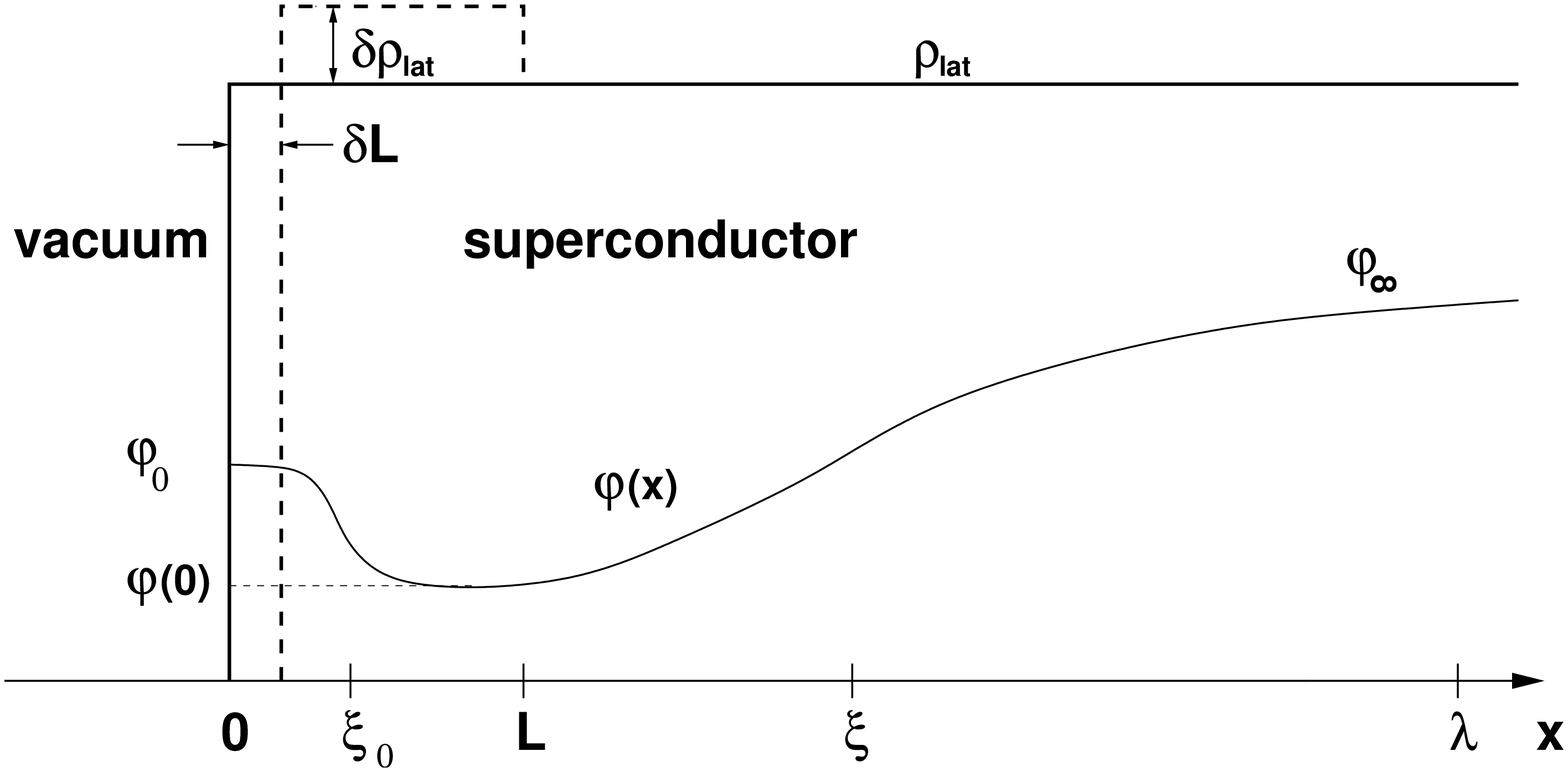,width=8cm}
\vskip 2pt
\caption{Schematic picture of the electrostatic potential at the 
surface. On the scale of the London penetration depth $\lambda$ 
the potential $\varphi(x)$ is concave, which corresponds to charge
accumulation. Below the GL coherence length $\xi$, the charge is
depleted leading to a convex potential. Within the BCS coherence 
length $\xi_0$, the potential makes a sharp step due to the surface 
dipole. $\varphi_0$ is the true surface potential, while 
$\varphi(0)$ is the extrapolated GL value. The dashed line indicates 
the virtual compression of the lattice.}
\label{f1}
\end{figure}
To introduce the surface dipole on an intuitive level we first 
assume that the system is close to the critical temperature. In 
this regime, the London penetration depth $\lambda$ and the 
GL coherence length $\xi$ are much larger than the BCS 
coherence length $\xi_0$. We can then define an intermediate 
scale $L$ such that $\xi_0\ll L\ll\xi,\lambda$, as sketched in 
Fig.~\ref{f1}. On the scale $L$, the GL wave function, the 
vector potential, and the electrostatic potential change only 
negligibly, e.g., $\varphi(x) \approx\varphi(x\to 0)\equiv
\varphi(0)$ for $\xi_0<x<L$. The extrapolation of the bulk potential 
towards the surface, $\varphi(0)$, has to be distinguished from 
the true surface potential $\varphi_0$. The difference 
$\varphi_0-\varphi(0)$ is caused by the surface dipole we aim 
to evaluate.

Close to the critical temperature we can follow the idea of Budd 
and Vannimenus \cite{BV73}. Let us assume a virtual compression of 
the crystal lattice such that the lattice charge density is 
removed from the surface layer of an infinitesimal width 
$\delta L$. The perturbation of the lattice charge density 
in the infinitesimal layer $0<x<\delta L$ is $\delta
\rho_{\rm lat}=-\rho_{\rm lat}$. The compression leads to an
increase of the charge density in the layer $\delta L<x<L$,
where $\delta\rho_{\rm lat}=\rho_{\rm lat}\delta L/L$, in 
accordance with the charge conservation.

Let us remind the basic idea of the Feynman-Hellmann theorem.
The lattice charge enters the jellium model of metals as an
external parameter. If one changes this external parameter, 
the situation corresponds to doing work on the system,
$\delta{\cal W}= S\int d x\delta\rho_{\rm lat}{\partial 
f\over\partial\rho_{\rm lat}}=S\int d x \delta\rho_{\rm lat} 
\varphi$, where $f$ is the density of the free energy including
the electrostatic interaction, and $S$ is the sample area. 
According to the Feynman-Hellmann theorem, the change of the 
electrostatic potential does not contribute to the work up to
the first order in $\delta\rho_{\rm lat}$. Now we can proceed
with the algebra. We split the integral into three parts. 
Since $\delta L$ 
is an infinitesimal displacement, the potential in the layer 
$0<x<\delta L$ can be replaced by the surface value 
$\varphi_0$. The surface region $\delta L<x<\xi_0$ gives a 
negligible contribution of the order of $\xi_0/L$. In the 
remaining bulk region $\xi_0<x<L$, the electrostatic potential 
is nearly constant and equals $\varphi(0)$. The work thus 
reads $\delta W= S \delta L \rho_{\rm lat} (\varphi(0)-\varphi_0)$. 

In the same time, the work increases the free energy ${\cal F}$ 
of the system, $\delta{\cal W}=\delta{\cal F}$. We denote the spatial
density of the electronic free energy by $f_{\rm el}$. Lower case 
letters denote free energies per unit volume, i.e., their 
densities, while the calligraphic upper case letters denote the 
corresponding total values. The total free energy changes as 
$\delta{\cal F}=-{\partial (f_{\rm el} S L) \over \partial (S L)} S
\delta L=\left(-f_{\rm el}+n{\partial f_{\rm el}\over\partial n} 
\right)S\delta L$. 
The first term results from the reduced volume, $SL\to S
(L-\delta L)$, and the second one from the corresponding 
increase of the electron density by $n\to n(1+\delta L/L)$.
We thus obtain a modification of the Budd-Vannimenus theorem,
\begin{eqnarray}
\rho_{\rm lat}\left(\varphi_0-\varphi(0)\right)&=&
f_{\rm el}-n{\partial f_{\rm el}\over\partial n}.
\label{e1}
\end{eqnarray}
This relation is our final result. It describes the step 
of the potential at the surface due to the surface dipole in 
terms of the free energy.

Formula (\ref{e1}) differs from the original Budd-Vannimenus 
theorem in three points. First, in the original Budd-Vannimenus
theorem, the surface potential is related to the potential 
$\varphi_\infty$ deep in the bulk. In (\ref{e1}), the 
extrapolation of the internal potential towards
the surface, $\varphi(0)$, appears instead. Second, in order
to cover systems at finite temperature, we use the free 
energy instead of the ground-state energy. Third, within the
original Budd-Vannimenus approach, the electron charge density
and the lattice charge density have to be equal because of the
charge neutrality. In our approach, the density of electronic 
charge differs locally from the lattice charge density, 
$en\ne -\rho_{\rm lat}$, due to the charge transfer on the 
scales $\xi$ and $\lambda$.

Let us demonstrate how the relation (\ref{e1}) can be used 
within the GL theory. To this end we introduce the GL free 
energy 
\begin{equation}
f_{\rm el}={1\over 2m^*}\left|\left(-i\hbar\nabla-e^*A\right)
\psi\right|^2+\alpha|\psi|^2+\beta|\psi|^4,
\label{e2}
\end{equation}
where $\psi$ is the GL wave function, $A$ is the vector 
potential and $m^*=2m$ and $e^*=2e$ are the mass and the 
charge of the Cooper pair. The GL parameters $\alpha$ and 
$\beta$ depend on the temperature and the electron density 
$n=n_n+2|\psi|^2$. Finally, we add the electromagnetic energy
\begin{equation}
f=f_{\rm el}+\varphi(\rho_{\rm lat}+en)-{\epsilon_0\over 2}E^2+
{1\over 2\mu_0}B^2,
\label{e3}
\end{equation}
with the magnetic field $B=\nabla\times A$ and the electric 
field $E=-\nabla\varphi$. 

Variations of the free energy with respect to its independent 
variables $A,\varphi,\psi,n_n$ yield the equations of motion
in Lagrange's form
\begin{equation}
-\nabla{\partial f\over\partial\nabla\nu}+{\partial f\over\partial
\nu}=0.
\label{e4}
\end{equation}
For $\nu=A$ the variational condition (\ref{e4}) yields the 
Ampere-Maxwell equation, for $\nu=\varphi$ the Poisson 
equation, for $\nu=\psi$ the GL equation, and for $\nu=n_n$ 
the condition of zero dissipation,
\begin{equation}
e\varphi=-{\partial f_{\rm el}\over\partial n}.
\label{e5}
\end{equation}
This condition allows one to evaluate the electrostatic potential 
in the bulk of the superconductor \cite{LKMB02}. Of course,  
one can add any constant to the electrostatic potential. The
actual choice (\ref{e5}) simplifies the relations we deal with below.
Another convenient choice would be to subtract the value of the 
potential in the non-magnetic state, i.e., deep inside the 
superconductor, so that $\varphi$ approaches zero in the bulk. 

Formula (\ref{e5}) does not cover the surface dipole on the
scale $\xi_0$, therefore at the surface it provides the 
extrapolated bulk value $\varphi(0)$. We can thus use (\ref{e5}) 
to rearrange the Budd-Vannimenus theorem (\ref{e1}) as
\begin{equation}
\rho_{\rm lat}\varphi_0=f_{\rm el}+\varphi(0)
\left(\rho_{\rm lat}+en\right).
\label{e6}
\end{equation}
Now all terms on the right hand side are explicit quantities 
which one obtains within the GL theory extended by the 
electrostatic interaction \cite{LKMB02}.

In customary GL treatments, the electrostatic potential and 
the corresponding charge transfer are omitted. This approximation 
works very well
for magnetic properties since the relative charge deviation, 
$\left(\rho_{\rm lat}+en\right)/\rho_{\rm lat}$, is typically 
of the order of $10^{-8}$ to $10^{-4}$ leading to comparably
small corrections in the GL equation. With the same accuracy 
one obtains the electronic free energy $f_{\rm el}$. Therefore 
it is possible to evaluate the surface potential using 
(\ref{e7}), which follows from (\ref{e6}) if terms proportional to 
$\left(\rho_{\rm lat}+
en\right)/\rho_{\rm lat}$ are neglected. 

So far we have restricted ourselves to systems close to the critical 
temperature when the validity conditions of the GL theory are well 
satisfied. In many cases, the GL theory is used beyond the limits 
of its nominal applicability, however. In these cases, the GL 
coherence length $\xi$ and/or the London penetration depth $\lambda$ 
are comparable to or even shorter than the BCS coherence length
$\xi_0$. Although the above derivation does not apply to this case,
it is possible to show that the Budd-Vannimenus theorem 
(\ref{e6}) provides physically consistent predictions in these 
cases, too.

We evaluate now the voltage which develops across a 
superconducting slab with the magnetic field parallel to the 
slab. For the slab geometry, the GL equation has an integral 
of motion, see Bardeen \cite{B55}. This integral can be 
obtained quite generally by the Legendre transformation of 
the free energy, $g=f-\sum_\nu(\partial f/\partial\nabla\nu)
\nabla\nu$. Indeed, if the fields $\nu$ obey the equations of 
motion (\ref{e4}), the gradient $\nabla g=0$ vanishes, i.e. 
$g={\rm const}$. With the help of
$\sum_\nu(\partial f/\partial\nabla\nu)\nabla\nu=
(\hbar^2/m^*)|\nabla\psi|^2-\epsilon_0E^2+B^2/\mu_0$, we 
can express the electronic free energy as
\begin{equation}
f_{\rm el}=g+{\hbar^2\over m^*}|\nabla\psi|^2-
{\epsilon_0E^2\over 2}
+{B^2\over 2\mu_0}-\varphi(\rho_{\rm lat}+en).
\label{e9}
\end{equation}
At the surface, the GL boundary condition demands that 
$\nabla\psi=0$ what implies $E=0$. The free energy on the 
surface thus reads
\begin{equation}
f_{\rm el}=g+{B^2\over 2\mu_0}-\varphi(0)
(\rho_{\rm lat}+en).
\label{e10}
\end{equation}
From the Budd-Vannimenus theorem (\ref{e6}) then follows 
$\rho_{\rm lat}\varphi_0=g+B^2/(2\mu_0)$, and since 
$g={\rm const}$, we obtain the potential difference across the 
surface as the difference of left and right magnetic pressures
\begin{equation}
\rho_{\rm lat} (\varphi_0^L-\varphi_0^R)=
{B_L^2-B_R^2\over 2 \mu_0}.
\label{e11}
\end{equation}
Exactly this value of the voltage has been observed by Morris 
and Brown \cite{MB71}.

As noticed already by Bok and Klein \cite{BK68}, there is a 
simple argument for the formula (\ref{e11}). The left hand side 
represents the electrostatic force (per unitary area) on the 
lattice, $F_{\rm elst}=\int_L^R dx\,E\rho_{\rm lat}=\rho_{\rm lat}
(\varphi_0^L-\varphi_0^R)$. The right hand side is the Lorentz 
force $F_{\rm Lor}=BJ$ with the mean magnetic field 
$B={1\over 2}(B_L+B_R)$ and the net current $J=\int_L^R dx\,j$ 
given by Ampere's rule, $B_L-B_R=\mu_0 J$. Since the 
electrostatic field provides the only mechanism by which the
force is passed from the electrons to the lattice, the two forces 
have to be equal, $F_{\rm Lor}=F_{\rm elst}$, which yields 
(\ref{e11}).

Turning the argument around, one can view the derivation of 
the Lorentz force as an alternative proof of formula (\ref{e6}). 
For the slab geometry, this is more general than the proof 
based on the Budd-Vannimenus approach, as the approach via the 
Lorentz force does not rely on the specific order of the 
characteristic lengths $\xi_0\ll\xi,\lambda$ needed above to 
define the intermediate length $L$. Moreover, the Legendre 
transformation provides the integral of motion for any 
modification of the GL free energy which is of the form 
$f_{\rm el}[\psi,\nabla\psi]$. 

On the other hand, within the Budd-Vannimenus approach the 
surface dipole is treated as a property of the superconducting 
condensate, what encourages us to hope that formula (\ref{e6}) 
or its approximation (\ref{e7}) can be used to obtain the surface 
potential also for cases when the magnetic field has a component 
perpendicular to the surface. In particular, we expect that it 
will be applicable also to the superconductors in the mixed state, 
especially to evaluate the electric field generated by vortices 
penetrating the surface \cite{B96}.

In conclusion, the Budd-Vannimenus theorem was modified to
the surface of a superconductor. It allows one to evaluate the 
electrostatic potential of the surface from the free energy and 
the bulk electrostatic potential near the surface. Within 
approximation (\ref{e7}) one obtains the surface potential from 
the free energy without the actual knowledge of the bulk potential. 

\medskip
This work was supported by M\v{S}MT program Kontakt ME601 
and GA\v{C}R 202000643, GAAV A1010312 grants. 
The European ESF program VORTEX is also acknowledged.

\end{document}